\documentclass[prb,aps,nobibnotes,twocolumn]{revtex4}
\usepackage{graphicx}%
\usepackage{dcolumn}
\usepackage{amsmath}   
\usepackage{bm}
\usepackage{dcolumn}
\usepackage{longtable}
\begin{document}

\title{\centering\Large\bf Reorganization energy of electron transfer 
                           at the solvent glass transition}
\author{Pradip K.\ Ghorai} 
\author{Dmitry V.\ Matyushov} 
\email[E-mail:]{dmitrym@asu.edu.}  
\affiliation{
  Department of Chemistry and Biochemistry and the Center for the
  Early Events in Photosynthesis, Arizona State University, PO Box
  871604, Tempe, AZ 85287-1604}
\date{\today}
\begin{abstract}
  We present a molecular-dynamics study of the solvent reorganization
  energy of electron transfer in supercooled water. We observe a
  sharp decrease of the reorganization energy at a temperature
  identified as the temperature of structural arrest due to cage
  effect as discussed by the mode coupling theory. Both the heat
  capacity and dielectric susceptibility of the pure water show sharp
  drops at about the same temperature. This temperature also marks the
  onset of the enhancement of translational diffusion relative to
  rotational relaxation signaling the break-down of the
  Stokes-Einstein relation.  The change in the reorganization energy
  at the transition temperature reflects the dynamical arrest of the
  slow, collective relaxation of the solvent related to Debye
  relaxation of the solvent dipolar polarization.
\end{abstract}
\maketitle

\section{Introduction}
\label{sec:1}
Theories of activated chemical dynamics and transport phenomena in
condensed phase are often based on transition-state ideas invoking
equilibrium thermodynamics to describe the reaction flux across the
transition-state surface separating the reactants from the products.
The Marcus-Hush theory of electron transfer (ET) reactions fully
relies on the transition-state formalism defining the ET activation
barrier in terms of two thermodynamic parameters, the free energy gap
and the nuclear reorganization energy. The former is the difference in
free energies between the final and initial ET states, and the latter
determines the curvature of two free energy parabolas.\cite{Marcus:93}
When the donor-acceptor energy gap $\Delta E$ is chosen as the reaction
coordinate, the classical reorganization energy is given through the
energy gap second cumulant:
\begin{equation}
  \label{eq:1-1}
  \lambda_s = \langle (\delta \Delta E)^2\rangle/2k_{\text{B}}T .
\end{equation}

The transition-state description becomes inapplicable when the time of
passage of the activation barrier is comparable to the relaxation time
of the condensed medium (solvent). The population of the activated
state gets depleted and one arrives at the friction-affected chemical
kinetics described by the Kramers theory\cite{Gardiner:97} and its
modifications.\cite{GroteHynes:80} For ET, this regime corresponds to
solvent-controlled reactions when the preexponential factor of the
rate is inversely proportional to a solvent relaxation
time.\cite{Barzykin:02} One can anticipate the next step in this
hierarchy of relaxation times when the time of the reaction itself
(not just the time of barrier passage) becomes comparable to the
solvent relaxation time. Such conditions, which apply to ultra-fast
reactions in high-temperature solvents and to reactions in
slowly-relaxing viscous solvents, will result in the loss of
ergodicity of the system and the break-down of the equilibrium
description of the reaction activation barrier (in contrast to the
alteration of the rate preexponent in the Kramers
description).\cite{DMjcp2:05} A general theory of chemical rates at
such conditions is still missing even though nonergodic behavior may
apply to a broad class of reactions in supercooled liquids and in
biopolymers.\cite{Dick:98} The latter case is particularly relevant to
the problem of nonergodic activation since the dynamics of
biopolymers is characterized by a broad spectrum of relaxation
times,\cite{Frauenfelder:88,Berg:05} and, for a given reaction rate,
at least a subset of nuclear modes may become nonergodic.

As the second cumulant of the difference in the solute-solvent
interaction potential (eq \ref{eq:1-1}), the solvent reorganization
energy of ET is related to a response function of the solvent to the
presence of the solute. The latter can be related to the
susceptibility of the pure solvent corresponding to a nuclear
collective coordinate coupled to the solute electronic states (dipolar
polarization and density for dipolar solvents, quadrupolar
polarization and density for non-dipolar solvents).\cite{DMjcp2:04}
Susceptibilities, which are second derivatives of a thermodynamic
potential, are known to show specific behavior at points of
thermodynamic instability (phase transitions) or at the onset of
nonergodicity (glass transition). It is of general interest to
understand how chemical reactions are affected by these special
points. For glass transitions, the heat capacity of a glassformer (the
second cumulant of energy at constant volume or the second cumulant of
entropy at constant pressure) passes through a peak which sharpens
with increasing fragility of the liquid.\cite{Angell:95} In this
Letter we present simulation results indicating that the solvent
reorganization energy of ET demonstrates a similar behavior.  It first
increases with lowering temperature and then sharply drops at the
temperature of kinetic glass transition.

\section{Simulation Results}
\label{sec:2}
We perform $NVE$ and $NPT$ molecular-dynamics (MD) simulations for a
system composed of one solute ($p$-nitroaniline) and $N=466$ water
molecules, with periodic boundary conditions. The extended simple
point charge model (SPC/E) is adopted for water. $p$-nitroaniline is a
small charge-transfer dye with a charge-transfer electronic transition
resulting in about 3.7 D change in the dipole
moment.\cite{Farztdinov:00} Its interaction with water is a sum of
Coulomb interactions between solute and solvent partial charges and
Lennard-Jones (LJ) potentials for which Lorentz-Berthelot combination rules
are used.  The $NVE$ simulations are done for a cubic simulation cell
with the side length of 24.075 \AA{} and water density of 0.997 gm/cm$^3$.
The cutoff radius of 12.0 \AA\ was employed.  Simulations have been
carried out in the range of temperatures from supercooled region (50
K) to superheated region (509 K).
\cite{hestanley00,pgdeben03,cvega05,Giovambattista:04} In parallel to
solvation simulations, simulations of the pure SPC/E water ($N=466$)
at the same thermodynamic conditions have been done.  Equilibration
has been carried out over a time period of 400 ps and production runs
have been accumulated over the duration of 1.0 ns at 509 K up to 6.0
ns at 50 K from configurations stored at the interval of 0.1 ps.

\begin{figure}[htbp]
  \centering
  \includegraphics*[width=6cm]{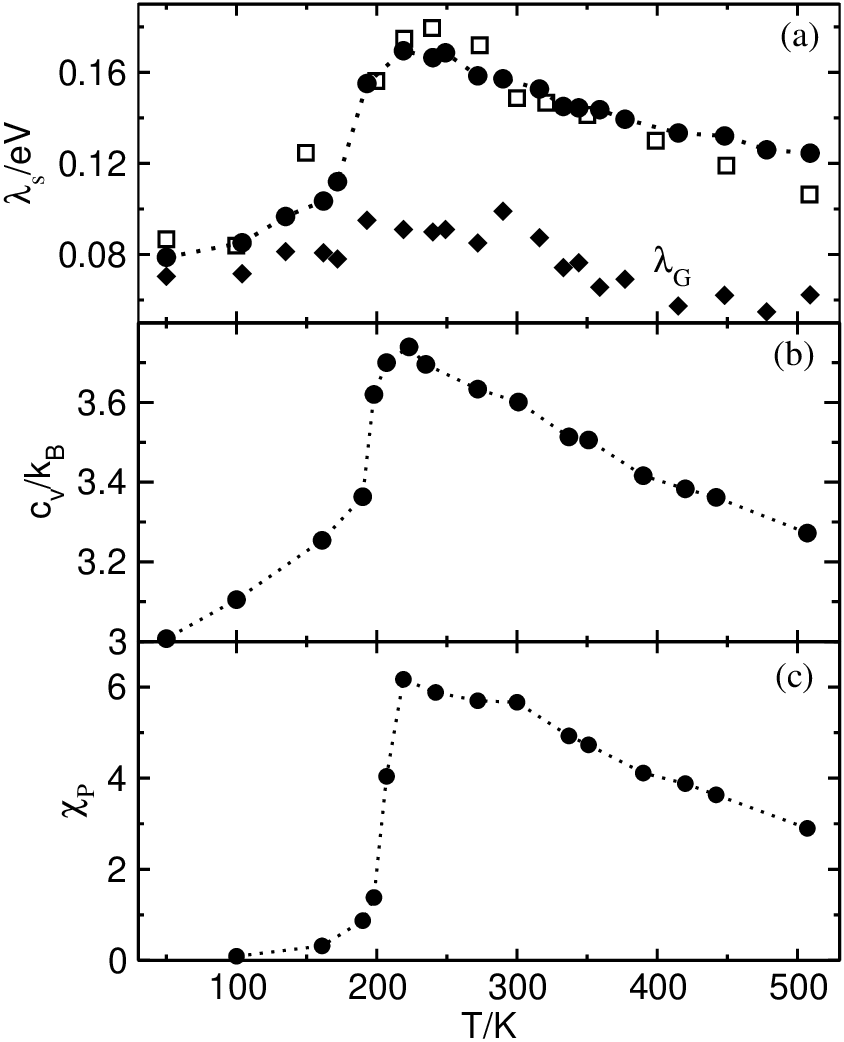}
  \caption{Reorganization energy (a), heat capacity (b), and dielectric susceptibility (c) 
    vs $T$. In (a), circles refer to $NVE$ simulations at $\rho = 0.9997$
    cm$^3$/g, squares indicate the $NPT$ simulations at $P=1$ atm,
    diamonds refer to the fast Gaussian component of the solvent
    reorganization energy, $\lambda_G$. Dotted lines are drawn to guide the eye. }
  \label{fig:1}
\end{figure}

Atomic charges of $p$-nitroaniline were obtained by fitting the
electrostatic potential from \textit{ab initio} electron structure
calculations using GAUSSIAN'03.\cite{gauss03} The ground state
geometry of $p$-nitroaniline was obtained on the MP2 level
(6-31+G$^*$) using X-ray data\cite{donohue56} for the initial
geometry. The excited-state charge distribution is obtained from SCI
calculations.  The ground, 7.18 D, and excited, 10.88 D, dipoles are in
reasonable agreement with the literature
data.\cite{Farztdinov:00,anrashid04} Coordinates and charges of
$p$-nitroaniline in the ground and excited states are listed in the
supplement.

Figure \ref{fig:1}(a) shows the solvent reorganization energy calculated
according to eq \ref{eq:1-1} in which the energy gap $\Delta E$ is replaced
with the difference of the solute-solvent interaction potential in the
charge-transfer ($S_4$) and ground ($S_0$) electronic
states.\cite{Farztdinov:00} The average is performed over the
simulation trajectories generated for the solute in the ground state.
Circles indicate constant-volume simulations and squares refer to
simulations at constant pressure of $P=1$ atm. The reorganization
energy increases with lowering temperature in the high-temperature
liquid\cite{DMjcp2:04} and then turns down and drops sharply at $T^* \simeq
219$ K.

\begin{figure}[htbp]
  \centering
  \includegraphics*[width=6cm]{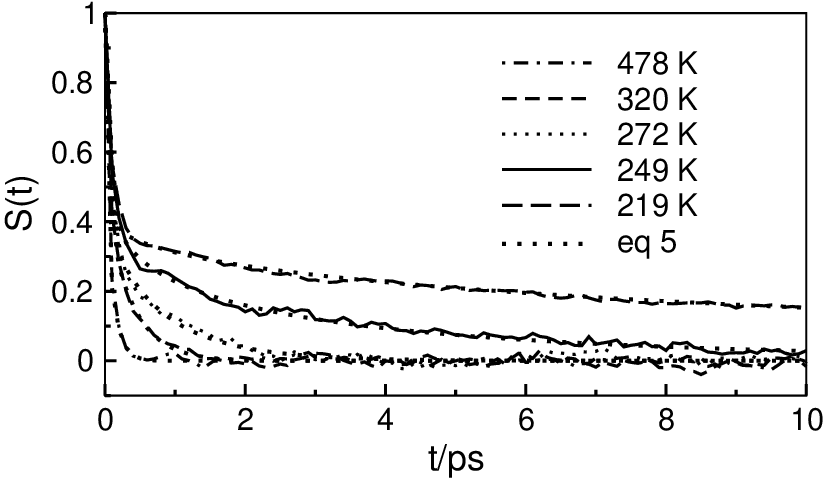}
  \caption{Stokes shift correlation function (eq \ref{eq:2-3}) of ground-state 
           $p$-nitroaniline in SPC/E water. }
  \label{fig:2}
\end{figure}

Shown in Figure \ref{fig:1}(b,c) are also the heat capacity,
\begin{equation}
  \label{eq:2-1}
     c_V/k_{\text{B}}= 3/2 + \langle(\delta E)^2\rangle/N(k_{\text{B}}T)^2   
\end{equation}
and dielectric susceptibility
\begin{equation}
  \label{eq:2-2}
  \chi_P = (k_{\text{B}}TV)^{-1} \langle (\delta \mathbf{M})^2\rangle  
\end{equation}
of SPC/E water ($E$ is the total energy and $\mathbf{M}$ is the total
dipole moment of a sample of liquid with volume $V$ containing $N$
molecules). Both solvent susceptibilities show the dependence on
temperature very similar to that of $\lambda_s$. Figure \ref{fig:1} thus
stresses the similarity in the temperature variation of the ET
reorganization energy and susceptibilities of the pure solvent at the
onset of nonergodicity at $T=T^*$.

The reorganization energy does not drop to zero, but instead levels
off at low temperatures. The low-temperature component of $\lambda_s$ is
related to fast phonon-like solvent modes which do not become
dynamically arrested below $T^*$. The fast reorganization component is
clearly seen as the initial Gaussian decay of the Stokes shift
correlation function\cite{Jimenez:94} (Figure \ref{fig:2})
\begin{equation}
  \label{eq:2-3}
  S(t) = C(t)/C(0),\quad C(t) = \langle\Delta E(t) \Delta E(0)\rangle. 
\end{equation}
$S(t)$ was fitted to a biphasic form containing the Gaussian (G) and
stretching exponential (E) parts
\begin{equation}
  \label{eq:2-4}
     S(t) = A_G e^{-(t/ \tau_G)^2} + (1 - A_G) e^{-(t/ \tau_E)^{\beta}} .
\end{equation}
The stretching exponent $\beta$ obtained from the fit is equal to one
above 272 K and starts to drop below this temperature reaching the
value of 0.34 at 219 K (see the supplement).
 
\begin{figure}[htbp]
  \centering
  \includegraphics*[width=6cm]{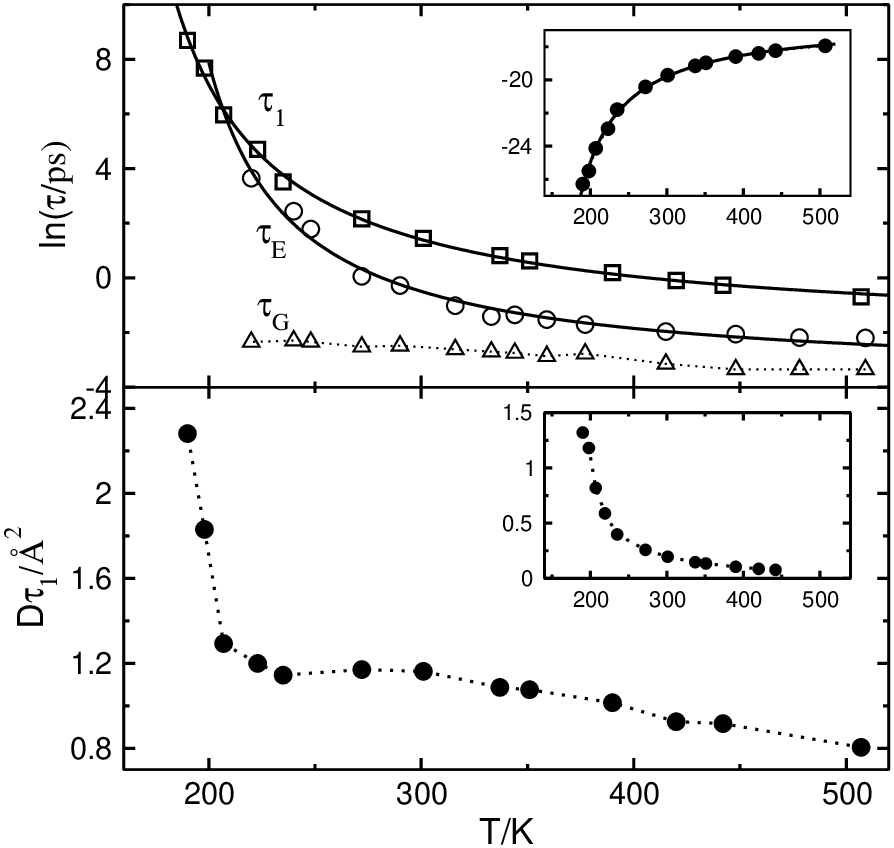}
  \caption{Upper panel: Relaxation times $\tau_G$, $\tau_E$ and $\tau_1$ vs $T$. 
    The inset shows natural logarithm of the diffusion coefficient $D$
    (m$^2$/s). The solid lines indicate the VF fits. Lower panel: The
    product of the self-diffusion coefficient and the rotational
    relaxation time $\tau_1$ vs $T$. The inset shows the maximum value
    $\alpha_2(t^*)$ of the non-Gaussian parameter $\alpha_2(t)$ defined by eq
    \ref{eq:2-7}.}
  \label{fig:3}
\end{figure}

From the fit of the simulated Stokes shift data, the Gaussian
relaxation time\cite{Jimenez:94} $\tau_G\simeq 70$ fs is weakly temperature dependent and is
approximately given by a linear function of $T$ (K): $\ln(\tau_G(T)/
\mathrm{ps})= -1.34 - 0.0042\times T$ (Figure \ref{fig:3}).  In contrast,
the exponential relaxation time $\tau_E$ increases sharply with lowering
$T$ and can be approximated by the Vogel-Fulcher (VF) law:
$\ln(\tau_E(T)/ \mathrm{ps}) = -3.73 + 447/(T-161)$, where $T$ is in
K.\cite{comT0} The Gaussian reorganization component, $\lambda_G = A_G\lambda_s$,
turns out to be weakly dependent on temperature (Figure \ref{fig:1}(a),
diamonds). The combination of the weak temperature dependence of $\lambda_G$
with the ultra-fast relaxation of the Gaussian component of $S(t)$
clearly indicates that it is the slow exponential component of $S(t)$
that becomes dynamically arrested at the transition to nonergodicity. The
drop of $\lambda_s$ at $T< T^*$ is thus equal to $\lambda_E = (1-A_G)\lambda_s$.

\begin{figure}[htbp]
  \centering \includegraphics*[width=6cm]{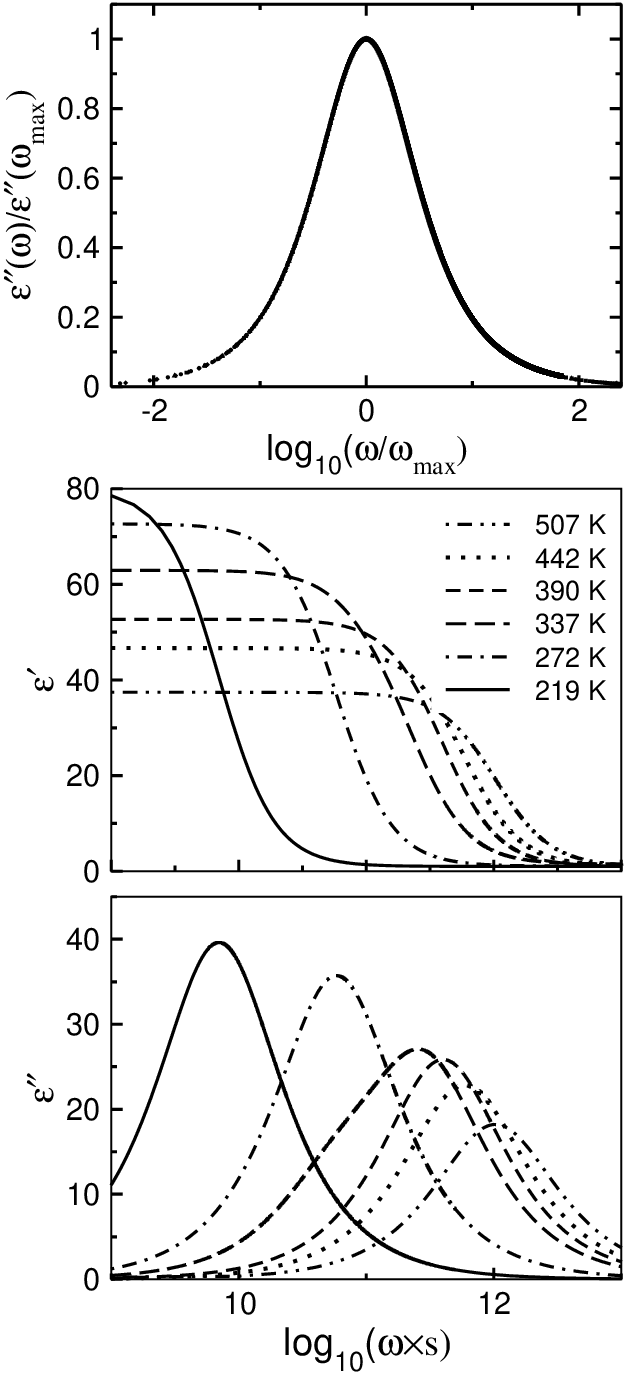}
  \caption{Real ($\epsilon'(\omega)$) and imaginary ($\epsilon''(\omega)$) parts of the frequency-dependent 
    dielectric constant of SPC/E water at different temperatures. The
    upper panel shows the re-scaled dielectric loss $\epsilon''(\omega)/
    \epsilon(\omega_{\text{max}})$ vs $\log_{10}(\omega/ \omega_{\text{max}})$ calculated at
    198, 219, 242, and 300 K; $\omega_{\text{max}}$ is the frequency
    maximum of the dielectric loss function. The data points at
    different temperatures cannot be distinguished on the plot scale.
  }
  \label{fig:4}
\end{figure}

The slow component of the Stokes shift dynamics can be identified with
collective orientational relaxation of the dipolar polarization of
water related to the frequency-dependent dielectric constant $\epsilon(\omega)$.
The variation of $\epsilon(\omega)$ with lowering temperature is characteristic of
many polar glassformers (Figure \ref{fig:4}).\cite{Dixon:90} In
particular, the peak $\omega_{\text{max}}$ of the dielectric loss $\epsilon''(\omega)$
shifts to lower frequencies with cooling the solvent. The width of the
dielectric spectrum, however, does not change, and the dielectric loss
data at different temperatures can be superimposed on one master curve
by proper re-scaling (Figure \ref{fig:4}, upper panel). The increase
in the width of dielectric loss is commonly associated with liquid
heterogeneity.\cite{Dixon:90} Figure \ref{fig:4} thus indicates that
dielectric response is homogeneous.
 
The collective relaxation of solvation shells reflected by $\tau_E(T)$
can be compared to the single-particle correlation function of the
dipole vector $\mathbf{m}(t)$:
\begin{equation}
  \label{eq:2-5}
  C_1(t) = \langle \mathbf{m}(t)\cdot\mathbf{m}(0)\rangle / \langle \mathbf{m}(0)^2 \rangle . 
\end{equation}
The relaxation time $\tau_1(T)$ for $C_1(t)$ shows a super-Arrhenius
temperature dependence similar to $\tau_E(T)$ (Figure \ref{fig:3}) which
can be approximated by the VF law: $\ln(\tau_1(T)/\mathrm{ps}) = -2.15 + 
578/(T-137)$.

What is the nature of the dynamical arrest occurring below $\simeq 219$ K?
Previous extensive simulations of SPC/E water have shown a power-law
temperature dependence of transport coefficients with the critical
temperature of ca.\ $T_c\simeq 186-200$
K.\cite{Gallo:96,Sciortino:96,Fabbian:99,Theis:00} Based on this
observation and scaling laws for the intermediate scattering function,
the critical temperature was assigned to the ideal
ergodic-to-nonergodic kinetic glass transition predicted by the
mode-coupling theory (MCT).\cite{Goetze:91} Our data for the
self-diffusion coefficient of SPC/E water $D(T)$ (obtained from MD
trajectories using the Einstein relation) embrace a broader range of
temperatures than those reported in refs \onlinecite{Gallo:96} and
\onlinecite{Sciortino:96}. The self-diffusion coefficient (Figure
\ref{fig:3}, inset) can be fitted by both the power-law ($\ln(D(T)\times \mathrm{s}/
\mathrm{m}^2)=-19+2.5\ln(T/181-1)$) and VF ($\ln(D(T)\times \mathrm{s}/
\mathrm{m}^2)=-16.5+540/(T-136)$) functions, with the latter providing
a better global fit. The limiting VF temperatures for single-particle
correlations reflected by $D(T)$ and $C_1(T)$ (136 K and 137 K,
respectively) are significantly lower than the corresponding
temperature (161 K) from the many-particle correlation time
$\tau_E(T)$.\cite{comT0}

The temperature $T^*$ at which $\lambda_s(T)$ and solvent susceptibilities
start to drop also marks the onset of the separation of translational
and rotational diffusion signaling the breakdown of the
Stokes-Einstein relation (Figure \ref{fig:3}, lower panel)
\begin{equation}
  \label{eq:2-6}
      D(T)\times \tau_1(T) \simeq \mathrm{const}    
\end{equation}
Indeed, the product of the diffusion coefficient and the rotational
relaxation time is approximately constant down to the temperature
$T_c\simeq 207$ K.\cite{comTc} If 165 K is adopted as the glass transition
temperature for water,\cite{comT0} the onset of translational
enhancement falls in the range 1.2--1.3 $T_g$ found in laboratory
experiment.\cite{Chang:97} This effect is commonly explained by either
static spatial\cite{Tarjus:95,Kivelson:98,Xia:01,Swallen:03} or
dynamic\cite{Jung:04} heterogeneity. In the frustration-limited domain
picture of Tarjus and Kivelson,\cite{Tarjus:95,Kivelson:98} the
turning temperatures in Figures \ref{fig:1} and \ref{fig:3} can be
associated with the onset of domain formation. This interpretation is
questionable, however, given the small size of our simulation box
which cannot incorporate mesoscopic domains. Note also that no
discontinuous change in pair distribution functions is observed at
$T^*$ except previously observed\cite{Sciortino:96,Donati:99}
sharpening of solvation shell peaks.

The results of our simulations better fit the picture of spatially
heterogeneous dynamics which assumes the presence in a supercooled
liquid of groups of mobile molecules.\cite{Jung:04,Schober:04,Stanley:04}
These groups may represent clusters of hydrogen bond defects in SPC/E
water,\cite{Stanley:04} chains of mobile particles in monoatomic
LJ fluids,\cite{Schober:04} or some other structures. The collective motion
of such mobile heterogeneities provides the enhancement of translational
diffusion which is correlated with the increase of the non-Gaussian
parameter\cite{Schober:04}
\begin{equation}
  \label{eq:2-7}
  \alpha_2(t) = 3\langle r^4(t)\rangle / 5\langle r^2(t)\rangle^2  - 1,  
\end{equation}
where $\langle r^2(t)\rangle $ is the mean square displacement. The maximum value
$\alpha_2(t^*)$ (Figure \ref{fig:3} lower panel inset) is reached at time
$t^*$.  Both $\alpha_2(t^*)$ and $t^*$ grow with cooling indicating
increasing dynamic heterogeneity of molecular translations. On the
other hand, dielectric loss data show no heterogeneity suggesting
a more homogeneous distribution of molecular rotations.

This interpretation is consistent with recent dielectric data by
Richert \textit{et al}.\cite{Richert:03} which do not show any
increase in spatial heterogeneity in the region of temperatures $T_g\leq
T\leq 1.2 T_g$ where previous reports\cite{Swallen:03} had indicated the
breakdown of the Stokes-Einstein relation. Although no broadly
accepted explanation of this phenomenon currently exists, the onset of
translation/rotation decoupling is normally associated with the
critical MCT temperature.  We will therefore resort to the
interpretation of temperature $T^*$ at which $\lambda_s$ dips to its
Gaussian component as the point of kinetic transition to nonergodicity
with many features of the critical temperature of an ideal glass
transition predicted by MCT.\cite{Goetze:91}

\section{Concluding remarks}
The MCT critical temperature $T_c$ located above the calorimetric
glass transition temperature marks the change in the mechanism of
relaxation in viscous liquids. Above $T_c$, transport phenomena and
response to a solute field are dominated by the rattling and
librations of particles in self-consistently maintained cages.  Below
$T_c$, on the other hand, the particles are almost arrested in the
free-energy landscape and transport and relaxation are triggered by
thermally activated hopping and rotation over saddle points. The
hopping mechanism is recorded by experiments with the observation
window up to 10$^2$ s in the temperature range $T_g\leq T \leq T_c$.  Many
redox reaction occur on the nanosecond and faster time-scale.  Since
hopping relaxation does not occur on the reaction time-scale, MCT
formalism may appear to be appropriate for describing ET reactions in
viscous supercooled liquids. The sharp drop of the solvent
reorganization energy close to $T_c$ found in our simulations is
therefore expected to mimic the laboratory ET experiment.

\begin{acknowledgments}
  This research was supported by the National Science Foundation
  (CHE-0304694). This is publication \#643 from the ASU Photosynthesis
  Center. Useful discussions with Prof.\ C.\ A.\ Angell and Prof.\ 
  R.\ Richert are gratefully acknowledged. 
\end{acknowledgments}  

\bibliographystyle{achemso}
\bibliography{reference,/home/dmitry/p/bib/chem_abbr,/home/dmitry/p/bib/photosynth,/home/dmitry/p/bib/et,/home/dmitry/p/bib/liquids,/home/dmitry/p/bib/solvation,/home/dmitry/p/bib/glass,/home/dmitry/p/bib/dm,/home/dmitry/p/bib/dynamics,/home/dmitry/p/bib/bioet}

\begin{thebibliography}{10}
\bibitem{comTc}
The temperature $T_c=207$ K from the plot of $D(T)\tau_1(T)$ vs $T$ is very
  close to the estimate for the critical temperature $T_c\simeq 208$ K from the
  molecular MCT with a restricted set of correlation functions (dipole
  approximation).\cite{Fabbian:99} This agreement is, however, fortuitous since
  the full version of the molecular MCT leads to a substantially higher
  temperature, $T_c\simeq 279$ K.\cite{Theis:00}.
\end{thebibliography}

\end{document}